\documentclass[prd,preprint,showpacs,preprintnumbers]{revtex4}

\usepackage{dcolumn}
\usepackage{bm}
\usepackage{epsf}

\begin{document}

\def\bu{B^+}
\def\bd{B^0_d} 
\def\bs{B^0_s}
\def\bsbar{\overline{B^0_s}}
\def\lb{\Lambda_b}
\def\bmix{B^0 \mbox{--} \overline{B^0}}
\def\bdmix{B_d^0 \mbox{--} \overline{B_d^0}}
\def\bsmix{B_s^0 \mbox{--} \overline{B_s^0}}
\def\dmd{\Delta m_d}
\def\dms{\Delta m_s}
\def\Zbb{Z^0 \rightarrow b\,{\overline b}}
\def\Zcc{Z^0 \rightarrow c\,{\overline c}}
\def\Zff{Z^0 \rightarrow f\,{\overline f}}
\def\Zqq{Z^0 \rightarrow q\,{\overline q}}
\def\Zuds{Z^0 \rightarrow u\,{\overline u},d\,{\overline d},s\,{\overline s}}

\begin{flushright}
SLAC-PUB-9285 \\
September 2002
\end{flushright}

\title{Search for time-dependent $B_s^0 - \overline{B_s^0}$ oscillations \\
using a vertex charge dipole technique}

\date{\today}

%
\def\iAOMORI{$^{(1)}$}
\def\iBRI{$^{(2)}$}
\def\iBRUN{$^{(3)}$}
\def\iBU{$^{(4)}$}
\def\iCOLO{$^{(5)}$}
\def\iCSU{$^{(6)}$}
\def\iFERR{$^{(7)}$}
\def\iFRAS{$^{(8)}$}
\def\iJHU{$^{(9)}$}
\def\iLBL{$^{(10)}$}
\def\iMASS{$^{(11)}$}
\def\iMISSI{$^{(12)}$}
\def\iMIT{$^{(13)}$}
\def\iMOSCOW{$^{(14)}$}
\def\iNAGO{$^{(15)}$}
\def\iOREG{$^{(16)}$}
\def\iOXF{$^{(17)}$}
\def\iPERU{$^{(18)}$}
\def\iRAL{$^{(19)}$}
\def\iRUTG{$^{(20)}$}
\def\iSLAC{$^{(21)}$}
\def\iSOONG{$^{(22)}$}
\def\iTENN{$^{(23)}$}
\def\iTOHO{$^{(24)}$}
\def\iUCSB{$^{(25)}$}
\def\iUCSC{$^{(26)}$}
\def\iVAND{$^{(27)}$}
\def\iWASH{$^{(28)}$}
\def\iWISC{$^{(29)}$}
\def\iYALE{$^{(30)}$}

\author{
  \baselineskip=.75\baselineskip  
\mbox{Kenji Abe\unskip,\iNAGO}
\mbox{Koya Abe\unskip,\iTOHO}
\mbox{T. Abe\unskip,\iSLAC}
\mbox{I. Adam\unskip,\iSLAC}
\mbox{H. Akimoto\unskip,\iSLAC}
\mbox{D. Aston\unskip,\iSLAC}
\mbox{K.G. Baird\unskip,\iMASS}
\mbox{C. Baltay\unskip,\iYALE}
\mbox{H.R. Band\unskip,\iWISC}
\mbox{T.L. Barklow\unskip,\iSLAC}
\mbox{J.M. Bauer\unskip,\iMISSI}
\mbox{G. Bellodi\unskip,\iOXF}
\mbox{R. Berger\unskip,\iSLAC}
\mbox{G. Blaylock\unskip,\iMASS}
\mbox{J.R. Bogart\unskip,\iSLAC}
\mbox{G.R. Bower\unskip,\iSLAC}
\mbox{J.E. Brau\unskip,\iOREG}
\mbox{M. Breidenbach\unskip,\iSLAC}
\mbox{W.M. Bugg\unskip,\iTENN}
\mbox{T.H. Burnett\unskip,\iWASH}
\mbox{P.N. Burrows\unskip,\iOXF}
\mbox{A. Calcaterra\unskip,\iFRAS}
\mbox{R. Cassell\unskip,\iSLAC}
\mbox{A. Chou\unskip,\iSLAC}
\mbox{H.O. Cohn\unskip,\iTENN}
\mbox{J.A. Coller\unskip,\iBU}
\mbox{M.R. Convery\unskip,\iSLAC}
\mbox{R.F. Cowan\unskip,\iMIT}
\mbox{G. Crawford\unskip,\iSLAC}
\mbox{C.J.S. Damerell\unskip,\iRAL}
\mbox{M. Daoudi\unskip,\iSLAC}
\mbox{N. de Groot\unskip,\iBRI}
\mbox{R. de Sangro\unskip,\iFRAS}
\mbox{D.N. Dong\unskip,\iSLAC}
\mbox{M. Doser\unskip,\iSLAC}
\mbox{R. Dubois\unskip,\iSLAC}
\mbox{I. Erofeeva\unskip,\iMOSCOW}
\mbox{V. Eschenburg\unskip,\iMISSI}
\mbox{S. Fahey\unskip,\iCOLO}
\mbox{D. Falciai\unskip,\iFRAS}
\mbox{J.P. Fernandez\unskip,\iUCSC}
\mbox{K. Flood\unskip,\iMASS}
\mbox{R. Frey\unskip,\iOREG}
\mbox{E.L. Hart\unskip,\iTENN}
\mbox{K. Hasuko\unskip,\iTOHO}
\mbox{S.S. Hertzbach\unskip,\iMASS}
\mbox{M.E. Huffer\unskip,\iSLAC}
\mbox{M. Iwasaki\unskip,\iOREG}
\mbox{D.J. Jackson\unskip,\iRAL}
\mbox{P. Jacques\unskip,\iRUTG}
\mbox{J.A. Jaros\unskip,\iSLAC}
\mbox{Z.Y. Jiang\unskip,\iSLAC}
\mbox{A.S. Johnson\unskip,\iSLAC}
\mbox{J.R. Johnson\unskip,\iWISC}
\mbox{R. Kajikawa\unskip,\iNAGO}
\mbox{M. Kalelkar\unskip,\iRUTG}
\mbox{H.J. Kang\unskip,\iRUTG}
\mbox{R.R. Kofler\unskip,\iMASS}
\mbox{R.S. Kroeger\unskip,\iMISSI}
\mbox{M. Langston\unskip,\iOREG}
\mbox{D.W.G. Leith\unskip,\iSLAC}
\mbox{V. Lia\unskip,\iMIT}
\mbox{C. Lin\unskip,\iMASS}
\mbox{G. Mancinelli\unskip,\iRUTG}
\mbox{S. Manly\unskip,\iYALE}
\mbox{G. Mantovani\unskip,\iPERU}
\mbox{T.W. Markiewicz\unskip,\iSLAC}
\mbox{T. Maruyama\unskip,\iSLAC}
\mbox{A.K. McKemey\unskip,\iBRUN}
\mbox{R. Messner\unskip,\iSLAC}
\mbox{K.C. Moffeit\unskip,\iSLAC}
\mbox{T.B. Moore\unskip,\iMASS}
\mbox{M. Morii\unskip,\iSLAC}
\mbox{D. Muller\unskip,\iSLAC}
\mbox{V. Murzin\unskip,\iMOSCOW}
\mbox{S. Narita\unskip,\iTOHO}
\mbox{U. Nauenberg\unskip,\iCOLO}
\mbox{H. Neal\unskip,\iYALE}
\mbox{G. Nesom\unskip,\iOXF}
\mbox{N. Oishi\unskip,\iNAGO}
\mbox{D. Onoprienko\unskip,\iTENN}
\mbox{R.S. Panvini\unskip,\iVAND}
\mbox{C.H. Park\unskip,\iSOONG}
\mbox{I. Peruzzi\unskip,\iFRAS}
\mbox{M. Piccolo\unskip,\iFRAS}
\mbox{L. Piemontese\unskip,\iFERR}
\mbox{R.J. Plano\unskip,\iRUTG}
\mbox{R. Prepost\unskip,\iWISC}
\mbox{C.Y. Prescott\unskip,\iSLAC}
\mbox{B.N. Ratcliff\unskip,\iSLAC}
\mbox{J. Reidy\unskip,\iMISSI}
\mbox{P.L. Reinertsen\unskip,\iUCSC}
\mbox{L.S. Rochester\unskip,\iSLAC}
\mbox{P.C. Rowson\unskip,\iSLAC}
\mbox{J.J. Russell\unskip,\iSLAC}
\mbox{O.H. Saxton\unskip,\iSLAC}
\mbox{T. Schalk\unskip,\iUCSC}
\mbox{B.A. Schumm\unskip,\iUCSC}
\mbox{J. Schwiening\unskip,\iSLAC}
\mbox{V.V. Serbo\unskip,\iSLAC}
\mbox{G. Shapiro\unskip,\iLBL}
\mbox{N.B. Sinev\unskip,\iOREG}
\mbox{J.A. Snyder\unskip,\iYALE}
\mbox{H. Staengle\unskip,\iMASS}
\mbox{A. Stahl\unskip,\iSLAC}
\mbox{P. Stamer\unskip,\iRUTG}
\mbox{H. Steiner\unskip,\iLBL}
\mbox{D. Su\unskip,\iSLAC}
\mbox{F. Suekane\unskip,\iTOHO}
\mbox{A. Sugiyama\unskip,\iNAGO}
\mbox{S. Suzuki\unskip,\iNAGO}
\mbox{M. Swartz\unskip,\iJHU}
\mbox{F.E. Taylor\unskip,\iMIT}
\mbox{J. Thom\unskip,\iSLAC}
\mbox{E. Torrence\unskip,\iMIT}
\mbox{T. Usher\unskip,\iSLAC}
\mbox{J. Va'vra\unskip,\iSLAC}
\mbox{R. Verdier\unskip,\iMIT}
\mbox{D.L. Wagner\unskip,\iCOLO}
\mbox{A.P. Waite\unskip,\iSLAC}
\mbox{S. Walston\unskip,\iOREG}
\mbox{A.W. Weidemann\unskip,\iTENN}
\mbox{E.R. Weiss\unskip,\iWASH}
\mbox{J.S. Whitaker\unskip,\iBU}
\mbox{S.H. Williams\unskip,\iSLAC}
\mbox{S. Willocq\unskip,\iMASS}
\mbox{R.J. Wilson\unskip,\iCSU}
\mbox{W.J. Wisniewski\unskip,\iSLAC}
\mbox{J.L. Wittlin\unskip,\iMASS}
\mbox{M. Woods\unskip,\iSLAC}
\mbox{T.R. Wright\unskip,\iWISC}
\mbox{R.K. Yamamoto\unskip,\iMIT}
\mbox{J. Yashima\unskip,\iTOHO}
\mbox{S.J. Yellin\unskip,\iUCSB}
\mbox{C.C. Young\unskip,\iSLAC}
\mbox{H. Yuta\unskip.\iAOMORI}
\it
  \vskip \baselineskip                   
  \centerline{(The SLD Collaboration)}   
  \vskip \baselineskip        
  \baselineskip=.75\baselineskip   
\iAOMORI
  Aomori University, Aomori , 030 Japan, \break
\iBRI
  University of Bristol, Bristol, United Kingdom, \break
\iBRUN
  Brunel University, Uxbridge, Middlesex, UB8 3PH United Kingdom, \break
\iBU
  Boston University, Boston, Massachusetts 02215, \break
\iCOLO
  University of Colorado, Boulder, Colorado 80309, \break
\iCSU
  Colorado State University, Ft. Collins, Colorado 80523, \break
\iFERR
  INFN Sezione di Ferrara and Universita di Ferrara, I-44100 Ferrara, Italy, \break
\iFRAS
  INFN Lab. Nazionali di Frascati, I-00044 Frascati, Italy, \break
\iJHU
  Johns Hopkins University,  Baltimore, Maryland 21218-2686, \break
\iLBL
  Lawrence Berkeley Laboratory, University of California, Berkeley, California 94720, \break
\iMASS
  University of Massachusetts, Amherst, Massachusetts 01003, \break
\iMISSI
  University of Mississippi, University, Mississippi 38677, \break
\iMIT
  Massachusetts Institute of Technology, Cambridge, Massachusetts 02139, \break
\iMOSCOW
  Institute of Nuclear Physics, Moscow State University, 119899, Moscow Russia, \break
\iNAGO
  Nagoya University, Chikusa-ku, Nagoya, 464 Japan, \break
\iOREG
  University of Oregon, Eugene, Oregon 97403, \break
\iOXF
  Oxford University, Oxford, OX1 3RH, United Kingdom, \break
\iPERU
  INFN Sezione di Perugia and Universita di Perugia, I-06100 Perugia, Italy, \break
\iRAL
  Rutherford Appleton Laboratory, Chilton, Didcot, Oxon OX11 0QX United Kingdom, \break
\iRUTG
  Rutgers University, Piscataway, New Jersey 08855, \break
\iSLAC
  Stanford Linear Accelerator Center, Stanford University, Stanford, California 94309, \break
\iSOONG
  Soongsil University, Seoul, Korea 156-743, \break
\iTENN
  University of Tennessee, Knoxville, Tennessee 37996, \break
\iTOHO
  Tohoku University, Sendai 980, Japan, \break
\iUCSB
  University of California at Santa Barbara, Santa Barbara, California 93106, \break
\iUCSC
  University of California at Santa Cruz, Santa Cruz, California 95064, \break
\iVAND
  Vanderbilt University, Nashville,Tennessee 37235, \break
\iWASH
  University of Washington, Seattle, Washington 98105, \break
\iWISC
  University of Wisconsin, Madison,Wisconsin 53706, \break
\iYALE
  Yale University, New Haven, Connecticut 06511. \break
\rm
} 
%


\begin{abstract}
We report a search for
$B_s^0 - \overline{B_s^0}$ oscillations using
a sample of $400\,000$ hadronic $Z^0$ decays collected
by the SLD experiment.
The analysis takes advantage of the electron beam polarization
as well as information from the hemisphere opposite
that of the reconstructed $B$ decay
to tag the $B$ production flavor.
The excellent resolution provided by the pixel CCD vertex detector
is exploited to cleanly reconstruct both $B$ and cascade $D$
decay vertices, and tag
the $B$ decay flavor from the charge difference between them.
We exclude the following values of the
$B_s^0 - \overline{B_s^0}$ oscillation frequency:
$\Delta m_s < 4.9$ ps$^{-1}$ and
$7.9 < \Delta m_s < 10.3$ ps$^{-1}$ at the 95\% confidence level.
\end{abstract}

\pacs{13.20.He, 13.25.Hw, 14.40.Nd}


\maketitle

\section{INTRODUCTION}

Neutral $B$ meson mixing plays a crucial role
in the determination of the Cabibbo-Kobayashi-Maskawa matrix
elements $V_{ts}$ and $V_{td}$.
However, the extraction of these elements from measurements
of the oscillation frequency is
complicated by the presence of hadronic uncertainties.
Lattice QCD calculations~\cite{Bernard} give a
15--20\% uncertainty in
the determination of $|V_{td}|$ from the precisely measured value
of the $\bd$ oscillation frequency, $\Delta m_d$,
due to the large uncertainty in the
decay constant and the ``B'' parameter for $\bd$ mesons.
This uncertainty can be reduced by extracting the ratio $|V_{ts}/V_{td}|$
from the ratio of $\bs$ and $\bd$ oscillation
frequencies, $\Delta m_s / \Delta m_d$, as many
theoretical uncertainties common to $\bs$ and $\bd$ mixing
cancel, leading to a reduced theoretical uncertainty of
5--10\%~\cite{Bernard,Kronfeld}.
In the framework of the Standard Model,
$|V_{ts}|$ is constrained by unitarity so the measurement
of the $\bs$ oscillation frequency will significantly improve
our knowledge of $|V_{td}|$.

  Measuring the oscillation frequency requires three ingredients:
the $B^0$ or $\overline{B^0}$ flavor at both production and decay,
and the proper decay time.
In the Standard Model, one expects $\Delta m_s$
to be an order of magnitude larger than $\Delta m_d$, making it
difficult to measure as the rapid oscillations have to be resolved
in the detector. The SLD pixel CCD vertex detector
is particularly well suited to this task thanks to its excellent
three-dimensional vertex resolution.
The analysis presented here determines the $B$ flavor at production
by exploiting the large forward-backward asymmetry of
polarized $\Zbb$ decays and uses additional information
from the hemisphere opposite that of the reconstructed $B$ decay.
The $B$ flavor at decay is tagged by the charge difference
between the $B$ and cascade $D$ decay vertices. 
This novel ``charge dipole'' technique relies heavily on the
high resolution of the vertex detector to
reconstruct separate secondary and tertiary vertices
originating from the $B \to D$ decay chain.
Throughout this paper, when reference is made to a specific state,
the charge conjugate state is also implied.

\section{APPARATUS}

The analysis uses a sample of $400\,000$ hadronic $Z^0$ decays collected
by the SLD experiment at the SLC between 1996 and 1998.
The detector elements most relevant to this analysis are the pixel
CCD Vertex Detector (VXD)~\cite{vxd3} for precise track position measurements
near the SLC Interaction Point (IP),
and the Central Drift Chamber (CDC)
for charged particle reconstruction and momentum measurement.
Charged tracks are reconstructed using hits in both VXD and CDC.
The track impact parameter resolution at high momentum
is measured to be 7.8~$\mu$m transverse to the beam direction ($xy$ plane)
and 9.7~$\mu$m in the plane containing the beam direction ($rz$ plane).
The centroid of the stable, micron-sized IP in the $xy$
plane is reconstructed using tracks in sets of
$\sim30$ sequential hadronic $Z^0$ decays,
with a measured precision of $\sigma_{IP} = 3.5\:\mu$m.
The $z$ coordinate of the IP is determined event by event using
the median $z$ position of the tracks at their point of closest
approach to the beam line.
A precision of $\sigma_z \simeq 17\:\mu$m is achieved for $\Zbb$ events,
as estimated from the Monte Carlo (MC) simulation. 

For a description of the SLD detector and
the MC simulation see Ref.~\cite{sldpaper}.
Decays of $B_s^0$ mesons are modeled according to the decay modes of
$B_d^0$ mesons~\cite{juliathesis}
assuming SU(3) flavor symmetry.
The simulation has been tuned to reproduce
tracking efficiencies and impact parameter resolutions measured in the data. 

\section{EVENT SELECTION}

The selection of $b$-hadron candidates proceeds in two stages.
First, hadronic $Z^0$ decays are selected.
Second, $\Zbb$ events are selected with an
inclusive topological reconstruction of $b$-hadron decays.
Vertex and kinematical information (mass and momentum) is used
at this stage to remove $udsc$ background. This second stage also
serves as a means to select tracks associated with the $b$-hadron decay
chain.

In the first stage, we select a sample of
$310\,488$ hadronic $Z^0$ decays.
The criteria, detailed in Ref.~\cite{sldpaper},
aim to remove leptonic final states, two-photon collisions and beam-induced
background events.
The remaining background, predominately due to $Z^0 \to \tau^+ \tau^-$ events,
is estimated to be $<0.1$\%. 

In the second stage,
each event is divided into two hemispheres with respect to the thrust axis.
In each hemisphere, a search is made for three-dimensional space points
with high track overlap density which are displaced from the IP,
taking the individual track resolution functions into account~\cite{zvtop}.
At this stage, no attempt is made to separate $B$ and $D$ decay
vertices and a single ``seed'' vertex (SV) is formed containing all
tracks from the $b$-hadron decay chain.
Tracks are required to have
$\ge3$ VXD hits, momentum transverse to the beam line
$p_{\perp}>250$ MeV/c, and
three-dimensional impact parameter $<3$ mm.
Tracks consistent with originating from a $\gamma$ conversion or from 
$K^0$ or $\Lambda$ decay are removed.
Identified vertices are required to be within a radius of 2.3 cm 
of the center of the beam pipe to remove vertices
resulting from interactions with the detector material.  
Two-prong vertices are required to have an invariant mass
at least 0.015 GeV/c$^2$ away from the nominal $K^0_s$ mass
to remove most of the remaining $K^0_s$ decays.
A set of two neural networks is used to suppress
light-flavor ($udsc$) background and to select the charged
tracks associated with the $b$-hadron decay~\cite{btom}.
The first neural network takes as input the distance between
the IP and the SV, that distance normalized by its error, and the
angle between the vertex axis and the total momentum vector of
the SV. 
The vertex axis is defined as the line joining the IP and the SV.
At least one good vertex is found in 72.7\% of bottom, 28.2\% of charm,
and 0.41\% of $uds$ quark hemispheres in the MC simulation. 
The second neural network improves the $b$-hadron charge and
mass reconstruction by associating tracks not used in the initial
vertexing to the seed vertex.
For each track, five parameters are taken as input to the neural
network. The first four 
(measured at the point of closest approach between the track
and the vertex axis) are the distance of closest approach,
the distance between the IP and the point of closest
approach along the vertex axis,
the ratio between that distance and the IP-to-SV distance,
and the angle between the track and the vertex axis.
The fifth parameter is the three-dimensional impact parameter
normalized by its error.
A $\sim$10\% inefficiency in full CDC-VXD track reconstruction
is partly recovered by also including tracks reconstructed
in the VXD alone (``VXD-only'' tracks).
Including VXD-only tracks with at least one hit in
each of the three VXD layers
reduces the tracking inefficiency to $\sim$3\% for heavy hadron decay products.
These tracks are used for vertex charge reconstruction
but not for vertex momentum or mass reconstruction
because of their poor momentum resolution.

A sample of $53\,709$ $\Zbb$ candidates is then selected by
requiring that either hemisphere in the event contain a seed
vertex with mass $M > 2$ GeV/c$^2$, where $M$ is partially corrected
for missing neutral decay products~\cite{sldpaper}.
The fraction of remaining light-flavor events is
estimated to be 2.9\% from the simulation.

\section{CHARGE DIPOLE RECONSTRUCTION AND DECAY FLAVOR TAG}

  In the last phase of the analysis, we aim to reconstruct
both $B$ and $D$ decay vertices in a given hemisphere.
We can then tag the $B^0$ or $\overline{B^0}$
decay flavor based on the charge difference between these
vertices. The charge difference is expected to be
$\pm 2$ (0) for decays with a charged (neutral) $D$ meson.
This charge dipole method relies on the fact that
the $B$ and cascade $D$ flight directions are very
nearly co-linear and the $B$ and $D$ decay points are separated
along the flight direction.
The final vertex reconstruction is carried out separately for each
event hemisphere and proceeds in two steps.

  In the first step, the $b$-hadron flight direction is determined using
a set of tracks selected by one of two different procedures.
The main track selection procedure uses the set of tracks
(including VXD-only tracks) associated with a seed vertex.
An alternative procedure is used if no SV is found in the hemisphere
(representing 15\% of the decays in the final sample).
In this case, tracks are required to have
either $\ge$3 VXD hits and $p_{\perp}>250$ MeV/c, or
$\ge$2 VXD hits, $\ge$23 CDC hits, $p_{\perp}>250$ MeV/c,
a $\chi^2/DOF < 8$ for both CDC and CDC+VXD track fits,
a first CDC hit with a radius $< 39.0$ cm,
a 2D impact parameter in $xy <$1 cm,
and a $z$ distance of closest approach to the IP $<1.5$ cm.
The latter procedure is useful to recover part of the
inefficiency of the SV finding algorithm for $b$ hadrons decaying close
to the IP, a region where the sensitivity to $\bs$ oscillations
is highest.
The tracks selected with either of the two above procedures are
then used to estimate the $b$-hadron flight direction.
This is done with a ``ghost'' track anchored
at the IP and initially given an error of 25 $\mu$m.
Each of the selected tracks is individually vertexed with the ghost
track and the sum of the two-prong vertex fit $\chi^2$ values is computed.
The direction of the ghost track is varied until this $\chi^2$ sum
is minimized---with the final direction
representing the best estimate of the $b$-hadron line of flight.
The ghost track error is then scaled such that
the largest single two-prong vertex fit $\chi^2$ equals 1.

  In the second step, the selected tracks are divided into subsets
defining decay vertices along the ghost track.
This is done using the following iterative procedure.
Initially, the probabilities for all two-track $+$ ghost track
and track $+$ IP candidate vertices are calculated.
The vertex with the highest fit probability is saved
and its track(s) is (are) removed from the list of available tracks.
If the highest probability vertex is one of the
track $+$ IP vertices, it defines a new IP vertex.
The highest probability vertex then serves as a new candidate
vertex for the next step, where again all combinations of the remaining
tracks with the new candidate vertex, with each other and the ghost track,
and with the IP are formed.
This process continues until all vertex fit probabilities
are lower than 1\%.
A single IP results from this procedure.
All remaining tracks (not incorporated in the IP or a multi-prong vertex)
are then combined with the ghost track to form one-prong vertices.
One advantage of this technique is that it allows for the reconstruction
of one-prong decay vertices, which are fairly common in both
$B$ and $D$ decays.
Details about the ability of this technique to reconstruct
the correct number of vertices can be found in Ref.~\cite{achou}.

Hemispheres with exactly two vertices (in addition to the IP) are selected.
The (secondary) vertex that is closer to the IP is labeled ``$B$''
and the (tertiary) vertex farther away is labeled ``$D$''.
If the $D$ vertex contains two or more tracks,
the $B$ vertex is further refined by adding a virtual
$D$ ``track'' to those tracks already attached to the $B$ vertex
and a new vertex fit is performed.
The $D$ track is formed using the $D$ vertex position and the
net momentum direction of the $D$ vertex tracks.
The error matrix of the $D$ track is scaled as a function of
the $D$ vertex mass to partially account for the fact that
the $D$ meson is not fully reconstructed.

The vertices must satisfy a series of criteria
to guarantee that the $B$ and $D$ vertices are well separated
and well reconstructed:
the $B$ vertex is downstream from the IP and
has a radius $< 2.2$ cm,
the distance between the $B$ and $D$ vertices ($L_{BD}$)
satisfies $250\:\mu$m $< L_{BD} < 1$ cm,
the $D$ vertex mass is smaller than 2.0 GeV/c$^2$
(assuming all tracks are pions),
the ghost track error is smaller than 300~$\mu$m, and
the cosine of the angle between the nearest jet axis direction
and the straight line connecting
the IP and the $B$ vertex is greater than 0.9.
For purposes of flavor tagging (see below),
the $B$ and $D$ vertex charges are required to be different.
Several cuts are applied to enhance the $B^0_s$ purity
and further improve the quality of the reconstruction for $B^0_s$ candidates:
the net charge $Q_{tot}$ of all tracks
associated with the decay chain is required to be zero and
neither vertex may contain a VXD-only track with  $p_\perp > 4$ GeV/c
to make sure that the charge is reliably reconstructed.
Finally, $\bd$ decays are suppressed by vetoing decays
including $D^{\ast +}$ candidates.
To this end, a veto is applied if slow pion candidates in either
$B$ or $D$ vertices have a $\pi D_{vtx} -D_{vtx}$ mass difference less
than 0.16 GeV/c$^2$.
This veto reduces the $\bs$ ($\bd$) selection efficiency by 5\% (13\%)
and results in a $\bs$ purity increase of 5\%.
For all data and MC events, we remove hemispheres already
containing a vertex selected in two complementary analyses,
one with fully reconstructed $D_s$~\cite{Dstracks} and the
other with a lepton + $D$ vertex.
Such a removal is necessary to avoid statistical correlations
and facilitate the combination of results
from different analyses.

\begin{figure}[tb]
  \vspace*{-10mm}
  \hspace*{6mm}
  \epsfxsize=11cm
  \epsfbox{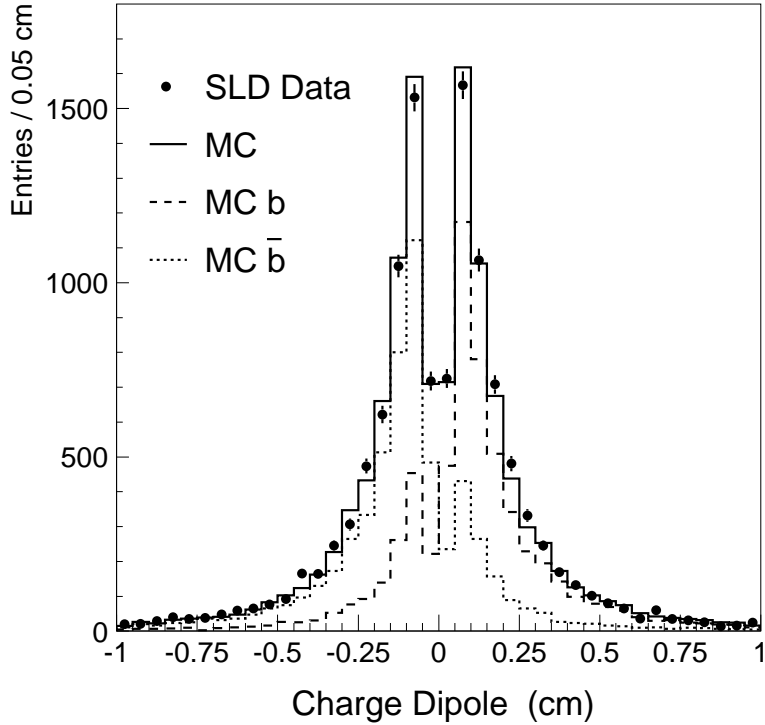}
  \vspace*{-5mm}
  \caption{\label{fig_dipo_distr}
  Distribution of the charge dipole for
  data (points) and Monte Carlo events (solid histogram).
  Also shown are the contributions from hadrons containing
  a $b$ quark (dashed histogram) or a $\bar{b}$ quark (dotted histogram).}
\end{figure}

  After applying all the above cuts, a sample of $11\,462$ decays remains.
The sample composition is
15.7\% $\bu$, 57.2\% $\bd$, 15.9\% $\bs$, 10.1\% $b$~baryon,
and 1.1\% $udsc$, as determined from the simulation.
The decay flavor is determined
by the sign of a vertex charge dipole defined as
$\delta Q \equiv L_{BD} \times SIGN (Q_D - Q_B)$,
where $Q_B$ ($Q_D$) is the charge of the $B$ ($D$) vertex.
Positive (negative) values of $\delta Q$ tag $\overline{B^0}$ ($B^0$)
decays.
Figure~\ref{fig_dipo_distr} shows the charge dipole distribution
for the data sample and also indicates the separation between
hadrons containing $b$ or $\bar{b}$ quarks in the simulation.
The mistag probability for selected $\bs$ ($\bd$) decays
is 22\% (24\%), as determined from the simulation.
The mistag probability depends on the charm content
of the decay products.
For example, the mistag probability is only 10\% for $B^0_s \to D_s X$ decays
($X$ represents any state without charm)
but 46\% for $\bs$ decays proceeding through a $b \to c \overline{c} s$
transition, where the latter category accounts for 24\% of the selected
$\bs$ decays.
Detailed studies of $B$ and $D$ vertex charge, multiplicity
and separation show that the simulation reproduces the data well
(see Ref.~\cite{juliathesis}).

  The proper time is computed using the measured decay length $L$
and momentum $p$ of the $\bs$ candidate, $T = L m_B / p c$.
The decay length is defined as the distance between the IP and the
$B$ vertex.
In the particular case of $B$ and $D$ vertices both containing
just one track (representing 17\% of the final sample),
the $B$ decay length is reconstructed by averaging
those decay lengths obtained from the ghost track + $B$ track vertex
and the $B$ track + $D$ track vertex.
The decay length residuals
for correctly (incorrectly) tagged $\bs \to D_{(s)} X$ decays
containing three or more tracks can be
described by the sum of a core Gaussian with a
width of $\sigma_{L} = 78~\mu$m ($109~\mu$m) 
and a tail Gaussian with a width of
$\sigma_{L} = 304~\mu$m ($584~\mu$m),
where the fraction of events in the core Gaussian is 60\%.
The momentum of the $\bs$ candidate is reconstructed
with an optimized average of two different methods.
The first method~\cite{bmoore} starts with tracks associated with the decay
and iteratively adds high energy ($E > 3$ GeV) neutral clusters
(from the electromagnetic calorimeter)
close to the $B$ vertex line of flight,
until the invariant mass of the decay products is close to the $B$ mass.
The second method~\cite{bdong} does not use calorimeter information but
relies on the associated tracks and the
amount of missing transverse momentum to estimate the $B$ momentum.
A parametrization with the sum of two Gaussians yields
a 60\% core width of $\sigma_{p}/p = 0.07$ and
a 40\% tail width of $\sigma_{p}/p = 0.21$ for
selected $\bs$ decays.

\section{PRODUCTION FLAVOR TAG}

\begin{figure}[tb]
  \vspace*{-10mm}
  \centering
  \epsfxsize11cm
  \leavevmode
  \epsffile{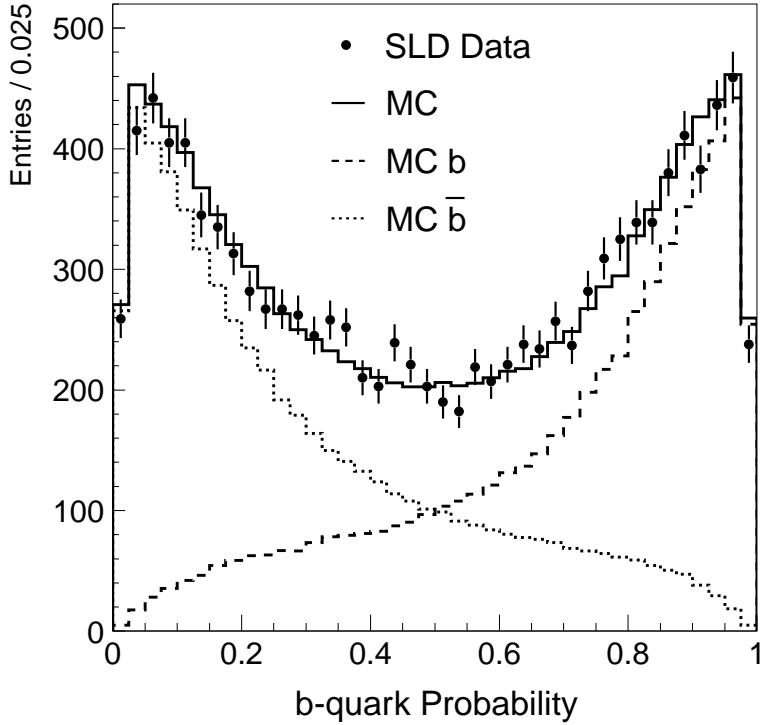}
  \vspace*{-5mm}
  \caption{ \label{fig_initag}
  Distribution of the computed $b$-quark probability at production for
  data (points) and Monte Carlo events (histograms) showing the $b$
  and $\bar{b}$ components (dashed and dotted histograms, respectively).}
\end{figure}

The $B$ flavor at production is determined with a combination of 
six tags: polarized forward-backward asymmetry, jet charge,
vertex charge, charge dipole, lepton and kaon tags,
where all but the first are combined in
an ``opposite hemisphere charge'' tag using a series of neural networks.
For a detailed description of the polarized forward-backward
asymmetry $\tilde{A}_{FB}$ see Ref.~\cite{sldpaper}. Briefly,
left- (right-) polarized electrons tag $b$ ($\bar{b}$) quarks
in the forward hemisphere, and $\bar{b}$ ($b$) quarks
in the backward hemisphere.
Averaged over the acceptance, this yields a mistag
probability of 28\% for our average electron beam polarization
of $P_e = 73\%$.
The opposite hemisphere charge tag employs tracks and topological
vertices reconstructed in the hemisphere opposite that of the
$\bs$ decay candidate. The available tags are:
(i) the jet charge tag, where tracks are used
to form a momentum-weighted track charge sum~\cite{sldpaper},
(ii) the total charge of tracks and charge dipole of a $b$-hadron decay,
(iii) the charge of a kaon identified in the Cherenkov Ring Imaging
Detector and associated with a $b$-hadron decay
(if more than one kaon is found, the total kaon charge is used),
and
(iv) the charge of a lepton originating from a $b$-hadron decay.
These tags are combined using a series of neural networks
to form an overall production flavor
tag characterized by a $b$-quark probability.
The neural networks take into account correlations between the
different tags as well as vertex charge, mass and decay length dependencies.
The corresponding average mistag rate is 29\%.
Finally, the result is analytically combined with the independent
$b$-quark probability from the $\tilde{A}_{FB}$ tag. The overall average
mistag probability is 22\% but the information is used on an
event-by-event basis, so that a significant fraction of the events have
a very high production flavor tag purity.
The combined tag is 100\% efficient.  
Figure~\ref{fig_initag} shows the $b$-quark probability distributions for data
and MC as selected in this analysis.
It indicates a clear separation between
$b$ and $\overline{b}$ quarks and good agreement between data and simulation. 

\section{FIT METHOD AND RESULTS}

Decays are tagged as mixed (unmixed) if
the production and decay flavor tags disagree (agree).
The probability for a decay to be in the mixed sample is expressed as
\begin{eqnarray}
  \lefteqn{{\cal P}_{mixed}(t,\dms) = \! f_u \frac{e^{-t/\tau_u}}{\tau_u}
      \left[w^P (1 - w^D_u) + (1 - w^P) w^D_u \right]} \nonumber \\
 & \! + & \!\!\frac{f_d}{2} \frac{e^{-t/\tau_d}}{\tau_d}
      \left(\sum_{k=1}^{5}~g_{dk}
      \left[(1-w^D_{dk}) (1 + [2 w^P - 1] \cos\dmd t)
           + w^D_{dk} (1 - [2 w^P - 1] \cos\dmd t) \right]
      \right) \nonumber \\
 & \! + & \!\!\frac{f_s}{2} \frac{e^{-t/\tau_s}}{\tau_s}
      \left(\sum_{k=1}^{5}~g_{sk}
      \left[(1-w^D_{sk}) (1 + [2 w^P - 1] \cos\dms t)
           + w^D_{sk} (1 - [2 w^P - 1] \cos\dms t) \right]
      \right) \nonumber \\
 & \! + & \!\! f_{bry} \frac{e^{-t/\tau_{bry}}}{\tau_{bry}}
      \left[w^P (1 - w^D_{bry}) + (1 - w^P) w^D_{bry} \right]
   \nonumber \\
 & \! + & \!\!\frac{f_{udsc}}{2}~F_{udsc}(t), \label{Pmix}
\end{eqnarray}
where $f_j$ represents the fraction of each $b$-hadron type and background
($j=u,d,s$, $bry$, and $udsc$ correspond to $\bu$, $\bd$, $\bs$, $b$~baryon,
and $udsc$ background),
$F_{udsc}(t)$ describes
the proper time distribution of the $udsc$ background,
$\tau_j$ is the lifetime for $b$ hadrons of type $j$,
$w^P$ is the production flavor mistag probability,
$w^D_u$ and $w^D_{bry}$ are the decay flavor mistag probabilities
for $\bu$ and $b$ baryons,
$w^D_{dk}$ and $w^D_{sk}$ are the decay flavor mistag probabilities
for $\bd$ and $\bs$ with the index $k= 1, ..., 5$
representing five different decay final states
($D^0 X$, $D^\pm X$, $D_s X$, charmed baryon $X$, and
final states resulting from a $b \to c \overline{c} s$ transition), and
$g_{dk}$ ($g_{sk}$) is the fraction of $\bd$ ($\bs$) decays into
each of the above final states.
A similar expression for the probability
${\cal P}_{unmixed}$ to observe
a decay tagged as unmixed is obtained by replacing the production flavor
mistag rate $w^P$ by $(1 - w^P)$.

  The quantities in Eq.~(\ref{Pmix}) are determined on
an event-by-event basis.
The fractions $f_j$ depend on the angle $\theta_T$
between the event thrust axis and the beam direction
to account for the decrease in the $Q_{tot}$ reconstruction
purity at high $|\cos\theta_T|$.
The mistag probability $w^P$ depends on $\cos\theta_T$, $P_e$,
and the opposite hemisphere charge neural network output.
The mistag probabilities $w^D_{jk}$ depend on the reconstructed
decay length in order to take into account the degradation of
the charge dipole tag close to the IP.
This effect is rather weak for $\bd$, $\bs$ and
$b$-baryon decays but it is significant for $\bu$ decays.
Finally, the decay final state fractions $g_{dk}$ and $g_{sk}$ are
parametrized as a function of $M$.

  Detector and vertex selection effects are introduced
by including a time-dependent efficiency function $\varepsilon (t)$
and convolving the above probability density functions
with a proper time resolution function ${\cal R}(T,t)$:
$  P_{mixed}(T,\dms) = \int_{0}^{\infty} {\cal P}_{mixed}(t,\dms)~{\cal R}(T,t)
  ~\varepsilon(t)~dt,
$
where $t$ is the ``true'' time and $T$ is the reconstructed time.
A similar expression applies to the unmixed probability $P_{unmixed}$.
Each term of the probability density function is divided by
a normalization factor given by
$\int_{0}^{10~\rm{ps}} (P_{mixed} + P_{unmixed})~dT$,
which accounts for the limited range of reconstructed proper time
considered in this analysis.
The resolution function is parametrized by the sum of two Gaussian and
two Novosibirsk functions~\cite{Novo},
\begin{eqnarray}
  {\cal R}(T,t)& = &
  0.36~f_{Gauss}(T,t;\sigma^{11},\mu_c)
+ 0.24~f_{Gauss}(T,t;\sigma^{12},\mu_c) \nonumber \\
  &  &
+ 0.24~f_{Novo}(T,t;\sigma^{21},\mu_t,\zeta)
+ 0.16~f_{Novo}(T,t;\sigma^{22},\mu_t,\zeta).
  \label{Resol}
\end{eqnarray}
The proper time resolution $\sigma^{ij}$ is a function
of proper time that depends on $p$, $\sigma_{p}$,
and $\sigma_L$,
\begin{equation}
  \sigma^{ij}(t) = \left[\left(\frac{\sigma_{L}^i\, m_B}{p\, c}\right)^2
  + \left(t\,\frac{\sigma_{p}^j}{p}\right)^2\right]^{1/2}~,
\end{equation}
where the index $i=1$ ($i=2$) corresponds to the core (tail) component of the
decay length resolution and the index $j=1$ ($j=2$)
corresponds to the core (tail) component of the relative momentum resolution.
Slight offsets in the decay length reconstruction are modeled by
the time-dependent parameters $\mu_c$ and $\mu_t$,
whereas an asymmetric tail is modeled by the time-dependent parameter
$\zeta$.
For each decay, the decay length resolution is computed from the vertex fit
and IP position measurement errors, with a scale factor determined using
the MC simulation (scale factors for correctly tagged decays
are typically $\sim$1.0 for the core and $\sim$2.0 for the tail components).
The momentum resolution is parametrized
as a function of the total track energy in each decay,
with parameters extracted from the MC simulation.
The efficiency $\varepsilon(t)$ is also parametrized using the
MC simulation.
All parametrizations are performed separately for each $b$-hadron type
and for right and wrong charge dipole tags.
As a consequence, different resolution functions are used for
final states containing one or two charm hadrons.

\begin{figure}[t]
  \vspace*{-10mm}
  \hspace*{6mm}
  \centering
  \epsfxsize=11cm
  \epsfbox{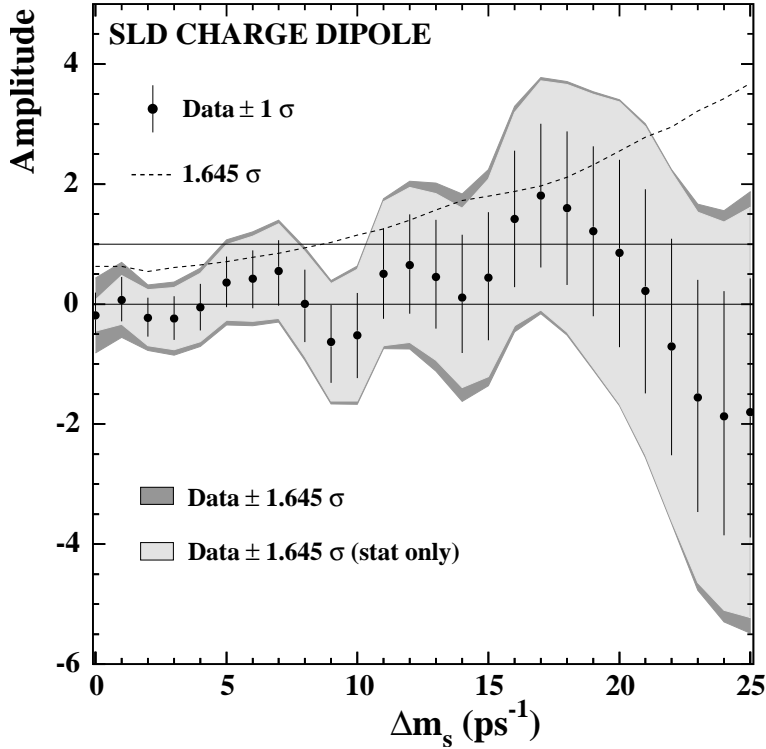}
  \vspace*{-4mm}
  \caption{\label{fig_afit}
  Measured $\bs$ oscillation amplitude as a function of $\dms$.
The light-grey (dark-grey) band is the 90\% confidence
level allowed region obtained using statistical (total) uncertainties.
Values of $\dms$ for which the band is below $A=1$ are
excluded at the one-sided 95\% confidence level.}
\end{figure}

  The study of the time dependence of $\bs$ mixing is carried out
using the amplitude method~\cite{Moser}.
A likelihood fit to the proper time distribution of mixed
and unmixed events is performed to determine the oscillation amplitude $A$
at fixed values of $\dms$. That is, in the expression for the mixed and unmixed
probabilities, one replaces $\left[ 1 \pm \cos(\Delta m_s t) \right]$
with $\left[ 1 \pm A \cos(\Delta m_s t) \right]$ and fits for $A$.
This method is similar to Fourier transform analysis
and has been tested extensively with simulated samples
generated at several different values of $\Delta m_s$.
The measured amplitude spectrum is shown in Fig.~\ref{fig_afit}.
The measured values are consistent with $A = 0$ for the whole range of
$\dms$ up to 25 ps$^{-1}$.
A signal for $\bs$ mixing, corresponding to an amplitude of $A=1$,
is ruled out for two regions of $\dms$ in the lower half of this range,
where $A + 1.645\,\sigma_A < 1$.

  As a check of the analysis,
we performed a fit for the $\bd$ oscillation frequency
and found a value of $\dmd = 0.537 \pm 0.030$(stat) ps$^{-1}$,
in agreement with the world average value of
$\dmd = 0.489 \pm 0.008$ ps$^{-1}$~\cite{PDG2002}.
Conversely, fitting for the $\bd$ oscillation amplitude at
a value of $\dmd = 0.489$ ps$^{-1}$ yields $A = 0.956 \pm 0.065$,
a value consistent with unity as expected for a signal.
Furthermore, the charge dipole mistag rate is measured directly from
the data for $\bd$ and $\bu$ decays and found to agree with MC values
within uncertainties (see next section).
To test the charge dipole tag for $\bs$ decays, we select
semileptonic decay candidates tagged as mixed and with $T < 2$ ps
to obtain a subsample with a $\bs$ purity of 28\%.
The charge dipole tag is found to agree with the
charge of the lepton in $(95 \pm 2)\%$ of the cases in the data,
to be compared with $(94 \pm 1)\%$ in the simulation.

\section{SYSTEMATICS}

  Systematic uncertainties have been computed following Ref.~\cite{Moser}
and are summarized in Table~\ref{tbl_syst} for several $\dms$ values.
The systematic uncertainty takes into account both the change in
the measured amplitude and in its error
\begin{equation}
  \sigma_A^{syst} = A^{var} - A^{nom}
   + (1 - A^{nom}) \frac{\sigma_A^{var} - \sigma_A^{nom}}{\sigma_A^{nom}},
\end{equation}
where $A^{nom}$ ($\sigma_A^{nom}$) is the nominal amplitude (error)
and $A^{var}$ ($\sigma_A^{var}$) is the amplitude (error) obtained
for a particular variation of the parameters of the likelihood
function.
It should be noted that the statistical
uncertainty dominates for all the $\dms$ values considered.
\begin{table}[t]
\caption{Measured values of the oscillation amplitude $A$ with a breakdown
   of the main systematic uncertainties for several $\dms$ values.}
\begin{center}
\begin{tabular}{lccc}
 $\dms$                  & ~~10 ps$^{-1}$  & ~~15 ps$^{-1}$ & ~~20 ps$^{-1}$ \\
 \hline
 \vspace{0.1cm}
 Measured amplitude $A$  &       $-0.517$  &       ~0.438   &       ~0.854  \\
 \vspace{0.1cm}
 $\sigma_A^{stat}$       &    $\pm 0.666$  &   $\pm 1.000$  &   $\pm 1.529$ \\
 \vspace{0.1cm}
 $\sigma_A^{syst}$       &
         $^{+0.217}_{-0.269}$ & $^{+0.444}_{-0.284}$ & $^{+0.254}_{-0.364}$ \\
 \hline
 \vspace{0.2cm}
 ${\cal B}(b \to \overline{\bs})$ &
         $^{-0.123}_{+0.162}$ & $^{-0.149}_{+0.136}$ & $^{-0.194}_{+0.170}$ \\
 \vspace{0.1cm}
 ${\cal B}(b \to b{\rm ~baryon})$ &
         $^{+0.028}_{-0.025}$ & $^{-0.004}_{-0.031}$ & $^{-0.041}_{-0.039}$ \\
 \vspace{0.1cm}
 $udsc$ fraction &
         $^{+0.002}_{+0.005}$ & $^{-0.023}_{-0.002}$ & $^{-0.055}_{-0.024}$ \\
 \vspace{0.1cm}
 Decay length resolution &
         $^{+0.031}_{-0.041}$ & $^{+0.003}_{-0.002}$ & $^{-0.022}_{+0.028}$ \\
 \vspace{0.1cm}
 Momentum resolution &
         $^{+0.042}_{-0.126}$ & $^{+0.024}_{-0.183}$ & $^{+0.056}_{+0.003}$ \\
 \vspace{0.1cm}
 Resolution function &
         $^{+0.115}_{-0.189}$ & $^{+0.416}_{-0.095}$ & $^{+0.099}_{-0.140}$ \\
 \vspace{0.1cm}
 Production flavor tag &
         $^{+0.018}_{-0.012}$ & $^{+0.026}_{-0.051}$ & $^{+0.042}_{-0.111}$ \\
 \vspace{0.1cm}
 Decay flavor tag &
         $^{+0.053}_{-0.050}$ & $^{+0.066}_{-0.092}$ & $^{+0.138}_{-0.206}$ \\
 \hline
\end{tabular}
\label{tbl_syst}
\end{center}
\end{table}

  Uncertainties in the sample composition are estimated by
varying the fraction of $udsc$ background by $\pm 20\%$
and the production fractions of $\bs$ and $b$~baryons according to
$0.100 \pm 0.012$ and $0.099 \pm 0.017$, respectively~\cite{WORKG}.
Since the $\bs$ purity is one of the most important parameters
in the analysis, we also varied the $\bd$ and $\bu$ branching
fractions to the five decay final states
($D^0 X$, $D^\pm X$, $D_s X$, charmed baryon $X$, and
$c \overline{c} s X$)
by twice the uncertainty given in Ref.~\cite{PDG2002}.
The corresponding branching fractions for
$\bs$ decays are varied by $\pm 20\%$ to account for
possible deviations due to SU(3) flavor breaking.
Furthermore, charge reconstruction uncertainties
are evaluated from comparisons between data and MC simulation,
especially relying on self-calibration tests detailed in Ref.~\cite{btom}.
To account for all uncertainties described above,
we vary the $\bs$ purity by $\pm 13.1\%$.
As a cross-check of the fraction of $\bs$ mesons in the selected sample,
we fit for $f_s$ with a $\chi^2$ fit to the
fraction of mixed decays as a function of proper time.
A value of $0.159 \pm 0.018$ is obtained, consistent with the MC
value of 0.159.
Other physics modeling uncertainties, which are due to
uncertainties in the $b$-hadron lifetimes and $\dmd$~\cite{WORKG},
are found to be negligible.

  Uncertainties in the modeling of the detector include
$\pm 7\%$ and $\pm 10\%$ variations in decay length
and momentum resolutions, respectively.
The decay length resolution uncertainty is determined from
3-prong $\tau$ decay vertices in the data and
the momentum resolution is confirmed to within 10\% by
comparing the width of the energy distribution in the data
with that expected from $b$-quark fragmentation~\cite{bdong}.
The systematic uncertainty due to the particular choice
of proper time resolution
function is evaluated by varying core fractions and offset
corrections.

  The production flavor tag systematic error accounts for the
uncertainties in the forward-backward asymmetry and
opposite hemisphere charge tags.
Dominant uncertainties for the forward-backward asymmetry
tag are the electron beam polarization $P_e = 0.73 \pm 0.02$
and the parity-violating parameter $A_b = 0.935 \pm 0.040$.
The uncertainty in the opposite hemisphere charge tag
is extracted from events in which both hemispheres have such
a tag. These double-tagged events provide a direct measurement
of the average mistag rate, which agrees very well
with the MC value and has an uncertainty of $0.008$.
The total uncertainty in the average production flavor
mistag probability is estimated to be $0.008$.
Furthermore, we allow for a deviation in the shape of
the $b$-quark probability distribution for the data.
The corresponding uncertainty is dominated by a
$7\%$ uncertainty in the slope of the $b$-quark
probability as a function of opposite hemisphere
charge neural network output, as extracted from
double-tagged events in the data.

  The decay flavor mistag probability is varied by $\pm 7\%$
for all $b$ hadrons.
This uncertainty is determined from $\chi^2$ fits to the
$\bd$ and $\bu$ mistag probabilities in the measured
dipole-tagged forward-backward asymmetry and the fraction
of mixed decays as a function of proper time.
These fits confirm that the mistag rate
for $\bd$ and $\bu$ decays is correctly modeled in the simulation.
To account for a possible deviation from SU(3) flavor
symmetry, an additional 7\% uncertainty is assigned to
$\bs$ decays. This value is obtained assuming
an additional 20\% uncertainty in the branching fraction
for $\bs$ decays proceeding via a $b \to c \bar{c} s$
transition (since these dominate the mistag rate).

  Including the systematic uncertainties,
the following ranges of $\bs$ oscillation frequencies are
excluded at the 95\% C.L.:
$\Delta m_s < 4.9$ ps$^{-1}$ and
$7.9 < \Delta m_s < 10.3$ ps$^{-1}$.
That is, the condition $A + 1.645\:\sigma_A < 1$ is satisfied for
those $\dms$ values.
The sensitivity to set a 95\% C.L. lower limit,
defined as the $\dms$ value below which $1.645\:\sigma_A < 1$,
is found to be 8.7 ps$^{-1}$.

\section{CONCLUSIONS}

  We have performed a search for $\bsmix$ oscillations
using a sample of $400\,000$ hadronic $Z^0$ decays collected by
the SLD experiment.
The cascade structure of $b$-hadron decays is reconstructed and
the $B$ flavor at decay is tagged with a novel charge dipole technique
that relies heavily on the excellent resolution of the CCD pixel
vertex detector.
The $B$ flavor at production is tagged using the polarized electron
beam as well as track and vertex information from the hemisphere
opposite that of the $\bs$ candidate.
With a final sample of $11\,462$ decays, we achieve
a sensitivity of 8.7 ps$^{-1}$ and exclude
the following values of the $\bs$ oscillation frequency:
$\Delta m_s < 4.9$ ps$^{-1}$ and
$7.9 < \Delta m_s < 10.3$ ps$^{-1}$
at the 95\% C.L.

  Combining these results with those obtained from a sample of
fully reconstructed $D_s$ mesons~\cite{Dstracks}
yields a combined SLD sensitivity of 11.1 ps$^{-1}$ and
excludes $\bs$ oscillation frequencies
$< 10.7$ ps$^{-1}$ at the 95\% C.L.
This result confirms previous studies performed
by the ALEPH, CDF, DELPHI, and OPAL Collaborations~\cite{BOSCWG},
and is among one of the most sensitive to date.

\section*{ACKNOWLEDGMENTS}
        We thank the personnel of the SLAC accelerator department and
the technical staffs of our collaborating institutions for their outstanding
efforts.  This work was supported by the Department of Energy, the National
Science Foundation, the Instituto Nazionale di Fisica of Italy, the
Japan-US Cooperative Research Project on High Energy Physics, and the
Particle Physics and Astronomy Research Council of the United Kingdom.

\end{document}